# A NOTE ON CORIOLIS QUANTUM STATES

G. DATTOLI, M. QUATTROMINI

ENEA - Dipartimento Tecnologie Fisiche e Nuovi Materiali
Centro Ricerche Frascati, Roma

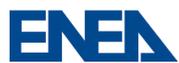

# A NOTE ON CORIOLIS QUANTUM STATES

G. DATTOLI, M. QUATTROMINI

ENEA - Dipartimento Tecnologie Fisiche e Nuovi Materiali
Centro Ricerche Frascati, Roma

# A NOTE ON CORIOLIS QUANTUM STATES

G. DATTOLI, M. QUATTROMINI

*Abstract*
*We introduce the Coriolis quantum states in analogy to the Landau states. We discuss their physical meaning and their role within the context of gravito-magnetic theory. We also analyse the experimental conditions under which they can be observed and their link with the Aharanov-Carmi effect.*

**Riassunto**
In questo lavoro si introducono gli stati quantistici di Coriolis in analogia a quanto si fa per gli stati di Landau. Ne discutiamo le proprietà e le condizioni per osservare tali stati per via sperimentale.

**Parole chiave:** Stati quantistici di Landau, Forze di Corolis, Teorie gravito-magnetiche, Effetto Aharanov-Bohm, Fase Aharanov-Carmi

**INDEX**

# A NOTE ON CORIOLIS QUANTUM STATES

In this note we introduce a set of quantum states associated with the Coriolis force.

They share some analogies with the Landau states [1], appearing in the study of quantum electron motion in a static magnetic field, and will be shown to be a mere consequence of the canonical commutation properties between momentum and position operators.

To make the analogy more transparent, we remind that, in some recent works [2-4], it has been proved that an operator like formalism, close to the method of quantum mechanics, is extremely useful to solve classical problems involving vector equations of the type

$$\partial_t \vec{R} = \vec{O} \times \vec{R},$$
$$\vec{R}(0) = \vec{R}_0 \tag{1.}$$

which occurs in many problems of classical and quantum mechanics regarding e. g. the evolution of the motion of classical particles under the influence of an external magnetic field [5] or of the Coriolis force [6].

The formal solution of eq. (1) can be written in terms of a kind of evolution operator, namely

$$\vec{R}(t) = e^{t\left[\overline{O\circ}\right]}\vec{R}_0 = \sum_{n=0}^{\infty}\frac{t^n}{n!}\left(\left[\overline{O}\right]\circ\right)^n\vec{R}_0 \tag{2},$$

where the successive applications of the operator $\left[\vec{O}\right]\circ$ to the initial vector $\vec{R}_0$ are specified by the following repeated applications of the vector product

$$\left(\left[\overline{O}\right]\circ\right)\vec{R}_0 = \overline{O}\times\vec{R}_0,$$

$$\left(\left[\overline{O}\right]\circ\right)^2\vec{R}_0 = \overline{O}\times(\overline{O}\times\vec{R}_0), \tag{3}.$$

$$\left(\left[\overline{O}\right]\circ\right)^3\vec{R}_0 = \overline{O}\times(\overline{O}\times(\overline{O}\times\vec{R}_0))...$$

Finally a closed expression can be obtained by taking advantage from the cyclical properties of the vector product[1], namely [2]

$$\vec{R}(t) = \cos(|\vec{O}|t)\vec{R}_0 + \sin(|\vec{O}|t)\,\vec{n}\times\vec{R}_0 + (1-\cos(|\vec{O}|t))(\vec{n}\cdot\vec{R}_0)\,\vec{n},$$

$$\vec{n} = \frac{\vec{O}}{|\vec{O}|} \tag{4}.$$

The solution given in eq. (4) is recognized as a rotation, written in a form known as Rodrigues' rotation [7].

The method we have outlined can be extended to the case in which the operator $\vec{O}$ is explicitly time dependent and not parallel to itself at different times. In this case the situation is mathematically equivalent to an often met problem in quantum mechanics, relevant to the solution of a Schroedinger equation ruled by an explicitly time dependent Hamiltonian operator, that does not commute with itself at different times. To this aim, ordering techniques, well known in Quantum mechanics, must be used to obtain the evolution of the vector $\vec{R}(t)$.

The use of quantum like[2] methods to solve problems of classical nature is extremely useful and the possibilities it offers should be studied carefully.

---

[1] By cyclical properties of the vector product we mean the triple vector product identity
$\vec{a}\times(\vec{b}\times\vec{c}) = (\vec{a}\cdot\vec{c})\vec{b} - (\vec{a}\cdot\vec{b})\vec{c}$.

[2] We mean the use of methods mathematically equivalent to those exploited in quantum mechanics.

However this is not the main point we want to touch on in this note. We have noted that the mathematical analogy between different problems leads to a solution which can exploited in different contexts, provided that the various quantities, appearing into the equation, are properly interpreted.

The point we want to rise is whether this analogy hides something more profound as a substantive physical equivalence.

The method we have outlined here applies in a quite natural way to the solution of the equation of motion of a body of mass $m$ with velocity $v$, moving in a rotating frame. The Coriolis force [5] acting on the body writes

$$\vec{F} = -2\,m\,\vec{\Omega} \times \vec{v} \tag{5}$$

where $\vec{\Omega}$ is the angular velocity vector, associated with the rotating frame.

The equations of motion yielding the time evolution of the velocity writes exactly as in (1), we could, therefore, exploit the evolution operator method to write its explicit time dependence in the form of a Rodrigues' rotation.

Equation (5) is also structurally equivalent to the Lorentz equation of motion and, at least from the mathematical point of view, the two problems are essentially the same.

To make the analogy more stringent we treat the solution of the Coriolis equation of motion using a real quantum treatment, we note, indeed, that the Hamiltonian of a rotating quantum system in a trapping potential can be written as

$$\hat{H} = \frac{1}{2\,m}(\vec{\hat{p}} - m\,\vec{\hat{\Gamma}})^2,$$

$$\vec{\hat{\Gamma}} \equiv (0, 2\,\Omega\,\hat{x}, 0), \tag{6a}$$

$$\vec{\Omega} = \vec{\nabla} \times \vec{\Gamma}$$

if the centrifugal part counterbalance the trapping potential. We have denoted by $\vec{\hat{p}}$ and $\vec{\hat{x}}$ the momentum and position operators, satisfying the canonical commutation brackets. The kinetic momenta

$$\hat{\Pi}_x = \hat{p}_x,$$

$$\hat{\Pi}_y = \hat{p}_y - 2\,m\,\Omega\,\hat{x} \tag{6b}$$

are covariant derivatives introduced on the basis of a kind of **minimal coupling,** in which the mass plays the role of the charge and $\vec{\Gamma}$ that of vector potential. The operators (6b) satisfy the rule of commutation

$$\left[\hat{\Pi}_x, \hat{\Pi}_y\right] = 2i\hbar m\Omega = i\frac{\hbar^2}{C_\Omega^2},$$

$$C_\Omega = \sqrt{\frac{\hbar}{2m\Omega}}$$

(7)

where $C_\Omega$ is a quantity, with the dimension of a length, which will be said **Coriolis Radius**, since it represents the counterpart of the **Cyclotron Radius** a key parameter in the theory of Landau States.

The operator $\hat{H}$ in eq. (6) takes the explicit form of a shifted harmonic oscillator Hamiltonian

$$\hat{H} = \frac{\hat{p}_x^2}{2m} + \frac{1}{2}m\tilde{\Omega}^2(\hat{x} - \hat{x}_\Omega)^2,$$

$$\hat{x}_\Omega = \frac{\hat{p}_y}{m\tilde{\Omega}}, \tilde{\Omega} = 2\Omega$$

(8).

The introduction of the creation annihilation operators

$$\hat{a}_c = \frac{C_\Omega}{\sqrt{2}\hbar}(\hat{\Pi}_x - i\hat{\Pi}_y),$$

$$\hat{a}_c^+ = \frac{C_\Omega}{\sqrt{2}\hbar}(\hat{\Pi}_x + i\hat{\Pi}_y)$$

(9),

satisfying the relation of commutation $\left[\hat{a}_c^+, \hat{a}_c\right] = \hat{1}$, allows to cast the operator (8) in the more concise form

$$\hat{H} = \hbar\tilde{\Omega}(\hat{a}_c^+\hat{a}_c + \frac{1}{2}),$$

(10).

The operator $\hat{p}_y$ commutes with the Hamiltonian and then it can be replaced by its eigenvalues $\hbar k_y$. The eigenstates and the eigenvalues of the above Hamiltonian are therefore

$$\Phi_n(x,y) = e^{i k_y y} \varphi_n(x - x_\Omega),$$

$$x_\Omega = \frac{\hbar k_y}{m \tilde{\Omega}},$$  (11)

$$E_n^\Omega = \hbar \tilde{\Omega} \left( n + \frac{1}{2} \right)$$

where $\varphi_n(x)$ are the harmonic oscillator Eigenfunctions and the Coriolis ground state wave function reads

$$\Phi_0(x,y) \propto e^{i k_y y} e^{-\frac{(x - x_\Omega)^2}{2 C_\Omega^2}}$$  (12).

Which is a gaussian packet, having the Coriolis Radius as rms width.

Let us now come to the moral of the previous discussion.

The derivation of the above states has been done in close analogy with the Landau states, regarding the quantization of the cyclotron orbit of charged particles in magnetic fields. We have just shown that they may exist, but before discussing the possibility of observing them experimentally it is worth wondering whether we have pointed out just a formal analogy or something deeper.

The so called gravito-magnetic theories [8] have stressed that effects like the Focault rotation can be associated with a vector potential of the type reported in eq. (5) and the relevant analysis has been conducted using the perspective of parallel transport and covariant derivatives [9]. The supposed existence of such a potential has given the opportunity of developing further speculations as those associated with a possible observation of the Ahronov-Bohm effect [10], involving the Coriolis vector potential. This phenomenon, suggested by Aharanov and Carmi (A-C) [11], has been observed experimentally using techniques of neutron interferometry [12].

We believe that the possibility of observing Coriolis quantum states could be an extremely interesting step within the framework of this phenomenology.

During the last years molecular rotation has provided interesting effects like electron spin rotation geometric phase shift or quantum decoherence, induced by stimulated Raman processes [see e.g. refs. 13]. It has been pointed that effects of the type discussed in this note, could be observed in rapidly rotating molecular systems [14].

Shen, He and Zhuang [14] have proposed the possibility of observing the A-C phase using the fullerene ($C_{60}$) molecule. Their analysis started with the observation that the valence electrons undergoing an inertial coupling to the molecular rapid rotation are ruled by the Hamiltonian

$$H(r) = \frac{1}{2 m_e} \left[ -i \hbar \vec{\nabla} - 2 m_e \vec{\Gamma} \right]^2 \qquad (13)$$

which is the same as our Hamiltonian (5).

The A-C phase associated with the above Hamiltonian is

$$\Delta \Phi = \frac{2 m_e \left| \vec{\Omega} \right|}{\hbar} \vec{n} \cdot \vec{A} =$$

$$= \frac{1}{C_{\Omega}^2} \vec{n} \cdot \vec{A} \qquad (14),$$

With $\vec{A}$ being the area vector circulated by the standing wave of the valence electrons on the $C_{60}$ molecular shell (see Fig. 1).

The detection of the A-C phase acquired by $C_{60}$ fullerene valence electrons, has the same level of difficulty of observing the Coriolis quantum states. By noting indeed that in high temperature orientationally disordered phase the angular frequency may be of the order of $10^{11} rad/s$ [15] and assuming for $\left| \vec{A} \right|$ a typical value of $3 \cdot 10^{-9} m^2$, we obtain for the A-C phase shift a value of about *1 mrad*. This effect will be responsible for a shift of the energy eigenvalues of the on shell energy electrons. Such a shift can be easily quantified as $\Delta E \cong 2 \hbar \Omega$ (see also refs. [14,16]), which is the same quantity yielding the energy separation between the Coriolis quantum states. This quantity may reach values of the order of a fraction of *meV,* in the orientationally disordered phase and could in principle be observed.

In conclusions, in this paper we have proved the existence of quantum states analogous to the Landau states. The analogy we have pointed may have deeper links with the gravito-magnetic effects like the A-C phase associated with the gravito-magnetic Lorentz field. We have also pointed out that these novel states are not at purely speculative level but amenable for an experimental study.

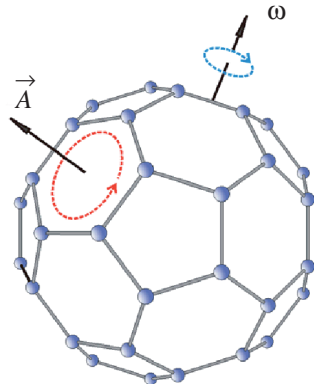

*Fig. 1 - Orientation and direction of the rotating vector and associated and of the associated Coriolis field in a*
$C_{60}$ *molecule*

In a forthcoming investigation we will enter more deeply into their nature and on the possible links with the quantum Hall effect.

## ACKNOWLEDGEMENTS

The Authors express their sincere appreciation to Dr. D. Babusci for the careful reading of the manuscript, for comments and suggestions.

# REFERENCES

[1]  R. Shankar "*Principles of Quantum Mechanics*", 2[nd] New York, Plenum Press (1994)

[2]  G. Dattoli, L. Mezi, and M. Migliorati, Il Nuovo Cimento **B117**, 781 (2002)

[3]  G. Dattoli, L. Mezi, and M. Migliorati, M., Il Nuovo Cimento **B118**, 493 (2003)

[4]  G. Dattoli, L. Mezi, and M. Migliorati, Il Nuovo Cimento **B119**, 565 (2004)

[5]  H. Goldstein, "*Classical Mechanics",* Addison Wesley Publishing Company (1965)

[6]  J.D. Jackson, "*Classical Electrodynamics*", John Wiley and Sons, New York (1975)

[7]  S.W. Sheppard, J. Guidance Control **1**, 223 (1978)
     O. Rodrigues, J. de Mathèmatique **5**, 380 (1840)

8.   H.O. Heaviside, The Electrician **31**, 81 (1893)
     S.J. Clark and R.W. Tucker, Class. Quantum Grav. **17**, 4125 (2000)

9.   M. Kugler, Am. J. Phys. **57**, 247 (1989)

10.  Y. Aharonov and D. Bohm, Phys. Rev. **115** , 485 (1959)

11.  Y. Aharonov and G. Carmi, Found. Phys. 3, 493 (1973)

12.  A.W. Overhauser and R. Colella, Phys. Rev. Lett. **33**, 1237 (1974)

13.  J.Q. Shen, S.L. He, Phys. Rev. **68 B**, 195421 (2003)
     Y.A. Serebrennikov, Phys. Rev. **73 B**, 195317 (2006)
     A. Adelsward and S. Wallentowitz, J. Opt. **6 B**, S147 (2004)

14.  J.Q. Shen, S. He and F. Zhuang, Eur. Phys. J. **33 D**, 35 (2005)

15.  R.D. Johnson et al., Science **255**, 1235 (1992)

16.  R.C. Haddon, L.E. Brus and K. Raghavachari, Chem. Phys. Lett. **125**, 465 (1986)